\documentstyle[12pt]{article}

\global\arraycolsep=1pt
\newcommand{\diag}{\mathop{\rm diag}}
\newcommand{\Tr}{\mathop{\rm Tr}}
\newcommand{\drawsquare}[2]{\hbox{%
\rule{#2pt}{#1pt}\hskip-#2pt
\rule{#1pt}{#2pt}\hskip-#1pt
\rule[#1pt]{#1pt}{#2pt}}\rule[#1pt]{#2pt}{#2pt}\hskip-#2pt
\rule{#2pt}{#1pt}}
\newcommand{\PSbox}[3]{\mbox{\rule{0in}{#3}\includegraphics{#1}\hspace{#2}}}

\newcommand{\Yfund}{\raisebox{-.5pt}{\drawsquare{6.5}{0.4}}}
\newcommand{\Ysymm}{\raisebox{-.5pt}{\drawsquare{6.5}{0.4}}\hskip-0.4pt%
        \raisebox{-.5pt}{\drawsquare{6.5}{0.4}}}
\newcommand{\Yasymm}{\raisebox{-3.5pt}{\drawsquare{6.5}{0.4}}\hskip-6.9pt%
        \raisebox{3pt}{\drawsquare{6.5}{0.4}}}

\newcounter{bean}

\begin{document}

\input epsf

\hfill MIT-CTP-2772

\hfill hep-th/9808026

\hfill August, 1998

\vspace{20pt}

\begin{center}
{\large {\bf NONPERTURBATIVE TESTS OF THE PARENT/ORBIFOLD CORRESPONDENCE
IN SUPERSYMMETRIC GAUGE THEORIES}}
\end{center}

\vspace{6pt}

\begin{center}
{\sl Joshua Erlich ~~and~~ Asad Naqvi}

{\it Center for Theoretical Physics,
Massachusetts Institute of Technology, Cambridge MA 02139}

{\tt jerlich@ctp.mit.edu, naqvi@ctp.mit.edu}

\end{center}

\vspace{8pt}

\begin{center}
{\bf Abstract}
\end{center}

\vspace{4pt}

It has been shown that a procedure analogous to orbifolding in string theory,
when applied to certain large $N$ field theories, leaves correlators
invariant perturbatively.  We test nonperturbative
agreement of some aspects of the orbifolded and non-orbifolded theories.  More
specifically, we find that the period matrices of parent and orbifolded
Seiberg-Witten theories are related, even away from the 't Hooft limit. We
also check that any large $N$ theory which has an infrared 
conformal fixed point and satisfies certain anomaly positivity constraints
required by theories with fixed points will continue to satisfy those 
constraints after orbifolding.  We discuss extensions of these results to 
finite $N$.

\vfill\eject

\section{Introduction}
Motivated by a correspondence between certain supergravity theories and
large $N$ conformal field theories \cite{maldacena}, and the preservation of 
this correspondence upon orbifolding of the supergravity theory 
\cite{kachru}, a relation between field theories and their orbifolds was
derived in \cite{bj}.  It was shown that at large $N$ all correlators of the 
orbifolded
theory are simply related in perturbation theory to the same correlators in
the parent theory.  Although we will continue to use the term ``orbifold''
interchangeably for this procedure acting on field theories and on a 
supergravity or string theory, for us orbifolds of field theories do not 
include
twisted sectors or anomalous U(1)'s.  Hence the motivation from the AdS
conjecture and the extension to orbifolds is tenuous.

In Sec. 2 we review orbifolding in field theories, and in Sec. 3 we
discuss the perturbative result of Bershadsky and Johansen.
In Sec. 4 we study nonperturbative extensions of this correspondence 
between field
theories and their orbifolds.  The result is that if a supersymmetric theory 
and its orbifold have Coulomb branches, the Seiberg-Witten period matrix of the
orbifolded theory is simply related to that of the parent theory.  This 
relation is valid for all $N$ and coupling $g$.  In Sec. 5 we study the 
anomaly positivity constraints \cite{josh} on 
supersymmetric theories with infrared fixed points and find that they are
satisfied in orbifolds of large $N$ theories with infrared fixed points.
We study two classes of theories at finite $N$ and find that the positivity
conditions hold for orbifolds of these theories, as well.  Conclusions are
summarized in Sec. 6.

\section{Orbifolding in Field Theory}

By orbifolding in field theory, we will mean removing from 
the theory all states which are not invariant under some discrete subgroup
 of the internal symmetry (gauge and global) of the theory (perhaps truncation
is a more appropriate term but we will continue to use orbifolding). Unlike
orbifolding in string theory, we  do not orbifold space-time so we
will look at quantum field theories in flat four dimensional Minkowski
space. In cases where the four dimensional theory can be realized as 
a world-volume theory on D-branes which are part of some 
brane configuration, orbifolding the space transverse to these
branes corresponds to orbifolding the field theory which lives on the
world volume of the branes except that we do not include the twisted
sector fields in the field theory orbifold
\cite{erich}. There are various restrictions on the type of orbifolds
allowed which come from string theory consistency requirements such as
tadpole cancelation.
 From the field theory point of view, the 
only restriction is from the requirement that the orbifolded
field theory does not have any gauge anomalies. We will now discuss some
examples of orbifolding in field theory. We will always use the 
regular representation (see below) of the orbifold group $G$ to embed 
it in the gauge group. If the field theory is realized as a world-volume
theory in some brane configuration, this requirement comes from the 
consistency of string theory. The importance of the regular representation is not clear from 
a purely field theoretic point of view. However, it simplifies the analysis
in perturbation theory \cite{bj}. We can also embed $G$ in the global symmetries 
of the theory. Different embeddings in the global symmetries will 
lead to different orbifold theories. 
\subsection*{SU($kN$) pure gauge theory orbifolded
by $Z_k$}
As discussed above, $Z_k$ is embedded by
its regular representation in the gauge group. In general, a discrete
group $G=\{g_1,g_2, \dots , g_k\}$ has a regular representation given by
$k\times k $ matrices $\gamma^a$ defined by $g_a g_i=g_j (\gamma^a)_{ji}$.
Using the fact that $g_a g_b\neq g_b$ unless $g_a=1$ (1 is the identity element
of the group which we will denote by $g_1$), we get
\[
\Tr \gamma^a =k \delta^a_1.
\]
It is easy to show using simple group representation theory that the 
regular representation is reducible and by a suitable change of basis can
be brought to a block diagonal form such that each irreducible representation
$R_i$ appears with multiplicity equal to its dimension $d_i=dim(R_i)$ along
the diagonal. 
This implies that $\sum_i d_i^2=k$. For the group $Z_k$, in an appropriate
basis, the regular representation matrices are given by 
\[
\gamma^a=\diag\{1,(\omega^a),(\omega^a)^2 \dots (\omega^a)^{k-1}\}, a\neq 1,
\]
\[
\gamma^1=\diag\{1,1,1, \dots,1\}, 
\]
where $\omega=e^{2 \pi i/k}$ and $\omega^k=1$. 
Now it is easy to embed the group $Z_k$ in the gauge group SU($kN$).
 The matrices, 
\[
\Gamma^a_N=\diag\{1,(\omega^a)\times {\mathbf 1}_N,(\omega^a)^2\times {\mathbf 1}_N \dots (\omega^a)^{k-1}\times {\mathbf 1}_N\}, a\neq 1,
\]
\[
\Gamma^1_N=\diag\{{\mathbf 1}_N,{\mathbf 1}_N,{\mathbf 1}_N, \dots, {\mathbf 1}_N\} 
\]
form a $Z_k$ subgroup of SU($kN$). 
This means that
the action of the orbifold group on the gauge field matrix $A_\mu=A_\mu^a T^a$
is given by $A_\mu \rightarrow \Gamma^i_N A_\mu \Gamma^{i^{\dagger}}_N$.
The components left invariant by the orbifold
group can then easily seen to be $N\times N$ blocks along the diagonals.
Hence the gauge group of the orbifolded
theory is SU($N)\times $SU($N)\times \dots \times $SU($N)$ ($k$ factors of
SU($N$)). Here, as mentioned in the introduction, we ignored anomalous gauge
U(1)'s.  
\subsection*{SU($kN$) theory with Adjoint scalars orbifolded
by $Z_k$}
Now consider an SU($kN$) gauge theory with complex scalars $\Phi$ in the
adjoint representation of the gauge group. This theory has a U($1$) global
symmetry 
$\Phi \rightarrow e^{i\alpha}\Phi$. We can embed $Z_k$
non-trivially in this global U(1) group as $\{\omega^j,j=0\dots k-1\}$ where
$\omega=e^{2\pi i /k}$. It is easy to check that the invariant scalars are
in $N \times N$ blocks shifted to the right of the diagonal. 
The matter content of the orbifolded theory is shown below (for $k=4$). 
\begin{center} 
\begin{tabular}{c c c c}
 SU$(N) $ &  SU($N)$ & SU($N)$ & SU($N)$   \\ 
$\Yfund $ &$ \overline{\Yfund}$  & 1& 1\\ 
$1 $ &$\Yfund $ &$\overline{\Yfund}$ & 1\\
$1 $ &$1 $ &\Yfund &$\overline{\Yfund}$ \\ 
$\overline{\Yfund} $ &$1 $ &1 &\Yfund \\ 
\end{tabular} 
\end{center}
\subsection*{${\cal N}=2$, SU($kN$) pure gauge theory orbifolded by $Z_k$}
In ${\cal N}=1$ language, the ${\cal N}=2$ pure gauge theory has a vector
superfield and a chiral superfield in the adjoint representation of 
the gauge group. This theory possesses a global U(1) symmetry under which
the gauge field and its fermionic partner transform as $(A_{\mu}, \lambda)
\rightarrow (A_{\mu},\lambda)$, and the adjoint scalar and its fermionic
partner as $(\phi,\psi) \rightarrow e^{i \alpha}(\phi, \psi)$. This symmetry
is anomalous but there is a discrete non-anomalous subgroup $Z_{2kN}$, which
in turn has a $Z_k$ subgroup generated by $\omega=e^{2\pi i/k}$. We identify
this $Z_k$ with the orbifold group. The gauge
group is embedded via the regular representation as usual. It is easy to see
that the orbifolded theory is an ${\cal N}=1$ supersymmetric SU($N)^{k}$
theory with chiral multiplets transforming as in the table above. We will
discuss the relation between the two theories in the Coulomb phase
 in section 4. 
\subsection*{${\cal N}=1$, SU($kN$) theory with $kF$ flavors orbifolded by $Z_k$}
This theory has a SU$(kF)_L \times $SU($kF)_R$ global symmetry and we
use $F$-fold copies of the regular representation to embed the orbifold group $Z_k$ in
each factor of the flavor group. We first embed the orbifold group trivially
in the other global symmetries (various U(1)'s). The orbifolded theory
is ${\cal N}=1$ supersymmetric SU($N)^k$ theory where each factor is 
disconnected and has $F$ flavors. We can also choose to embed the 
orbifold non-trivially in the U($1)_R$ symmetry under which the gauginos
and flavors have charge +1 (implying that the fermionic
quarks are uncharged)(this theory is discussed in detail in \cite{martin}). This symmetry is anomalous but has a non-anomalous $Z_{kN}$ subgroup which in turn has a $Z_k$ subgroup. This we identify with the orbifold group. 
We will use $Q$,$\bar{Q}$, $\Psi$,$\bar{\Psi}$ for
the scalar and fermionic components of the superfields transforming in the
fundamental and anti-fundamental representation of the gauge group. 
Then, under the orbifold group, the fields will transform as
$A_{\mu} \rightarrow \Gamma_N^a A_\mu \Gamma_N^{a^{\dagger}}$, 
$\lambda \rightarrow \omega^{a-1} \Gamma_N^a \lambda \Gamma_N^{a^{\dagger}}$, 
$Q \rightarrow {\omega}^{a-1} \Gamma_N^a Q \Gamma_F^{a^{\dagger}}$, 
$\bar{Q} \rightarrow{\omega^*}^{a-1} \Gamma_N^{a^{\dagger}} \bar{Q} \Gamma_F^{a}$,
$\Psi \rightarrow  \Gamma_N^a \Psi \Gamma_F^{a^{\dagger}}$, 
and $\bar{\Psi} \rightarrow   \Gamma_N^{a^{\dagger}} \bar{\Psi} \Gamma_F^{a}$, where $\omega=e^{2\pi i /k}$. The orbifolded theory has no supersymmetry and the following matter content (for $k=3$).
\begin{center} 
\begin{tabular}{c|c c c c c c c c c }\
& SU$(N) $ &  SU($N)$ & SU($N)$ & SU($F)_L$ & SU($F)_L$ & SU($F)_L$ &SU($F)_R$ & SU($F)_R$ & SU($F)_R$   \\ \hline
$\lambda_1$ &$\Yfund $ &$ \overline{\Yfund}$  & 1& 1 &1 &1 &1&1&1\\ 
$\lambda_2 $& $1 $ &$\Yfund $ &$\overline{\Yfund}$ & 1 & 1 & 1&1&1&1\\
$\lambda_3$ & $\overline{\Yfund} $ &1 &\Yfund & 1 & 1 & 1 &1&1&1\\
$Q_1$ & \Yfund &1 & 1 &1&$\overline{\Yfund}$  & 1 & 1&1&1 \\
$Q_2$ &1& \Yfund & 1 &1 & 1& $\overline{\Yfund}$  & 1&1&1 \\
$Q_3$ &1&1& \Yfund & $\overline{\Yfund}$&1&1&1&1&1 \\
$\bar{Q}_1$ &$\overline{\Yfund}$ &1 & 1 &1&1  & 1 & 1&1&$\Yfund$ \\
$\bar{Q}_2$ &1& $\overline{\Yfund}$ & 1 &1 & 1& 1  & \Yfund&1&1 \\
$\bar{Q}_3$ &1&1& $\overline{\Yfund}$ & 1&1&1&1&\Yfund&1 \\  
$\Psi_1$ & \Yfund &1 & 1 &$\overline{\Yfund}$  & 1 & 1&1&1&1 \\
$\Psi_1$ &1& \Yfund &1 & 1 &$\overline{\Yfund}$  & 1 & 1&1&1 \\
$\Psi_1$ &1&1&  \Yfund &1 & 1 &$\overline{\Yfund}$  & 1 & 1&1 \\
$\bar{\Psi}_1$ & $\overline{\Yfund}$ &1 & 1 &\Yfund  & 1 & 1&1&1&1 \\
$\bar{\Psi}_2$ &1& $\overline{\Yfund}$ &1 & 1 &\Yfund  & 1 & 1&1&1 \\
$\bar{\Psi}_3$ &1&1& $\overline{\Yfund}$ &1 & 1 &\Yfund  & 1 & 1&1 \\
\end{tabular} 
\end{center}

\subsection*{Orbifolding $\Ysymm$ and $\Yasymm$ under SU($kN$) by $Z_k$}
The symmetric tensor may transform under some global U(1) as $\Ysymm$  
$(\Yasymm)\rightarrow e^{i \alpha} \Ysymm$ 
$(\Yasymm)$. If we embed the
orbifold group trivially in this U(1), $\Ysymm$ $(\Yasymm)$ under SU($kN$) will become
$\Ysymm$ ($\Yasymm$) under each of the factors of SU($N)^k$. However, if we choose to
embed it non-trivially as $\{\omega^j$, $j=0 \dots k-1\}$, we get bifundamentals
under the $k$ factors if $k$ is odd and bifundamentals under $k-1$ factors
and $\Ysymm$ ($\Yasymm$) under one of the factors if $k$ is even. 

\section{Perturbative Correspondence between Parent and Orbifolded Theories}
In this section, we review some of the arguments by Bershadsky and Johansen
\cite{bj} who proved that at large $N$, all correlators of the orbifolded theory are 
identical to the corresponding correlators in the parent theory upto a
rescaling of the coupling constants. In the orbifolding procedure, the
coupling constants $g_{orb}$ of the various factors of the product groups are 
the same as that of the parent theory. However, if we define $g_{orb}^2=k g_{parent}^2$, the correlators of the two theories are identical.  

We define a projector onto states which are invariant under the orbifold
group by 
\begin{equation}
P=\frac{1}{k}\sum_{a=1}^{k} r^a, 
\end{equation}
where $r^a$ are matrices in a particular (generally reducible) representation
of the orbifold group. It is easy to show that $P^2=P$ and $P=1$ in
the trivial representation and that $P=0$ in all other irreducible 
representations. Thus when acting of some field in representation $R$, 
the projector projects the invariant components. For example, for adjoint
fields,  the projector can be defined as
\begin{equation}
P=\frac{1}{k}\sum_{a=1}^k r_a \otimes \Gamma_a^{\dagger} \otimes \Gamma_a
\end{equation}
where $r_a$ is the action of the $Z_k$ subgroup of the global symmetry, 
$\Gamma_a^{\dagger}$ acts on the anti-fundamental index and $\Gamma_a$ acts
on the fundamental index. At large $N$, the perturbation series is dominated
by planar Feynman diagrams with arbitrary number of loops\cite{tHooft}. Each diagram
factorizes into product of certain kinematic (group theory) factor  and
another factor with  complicated momentum and spin dependence which is 
independent of the internal symmetry structure of the diagram.
\begin{figure}\begin{center}
\PSbox{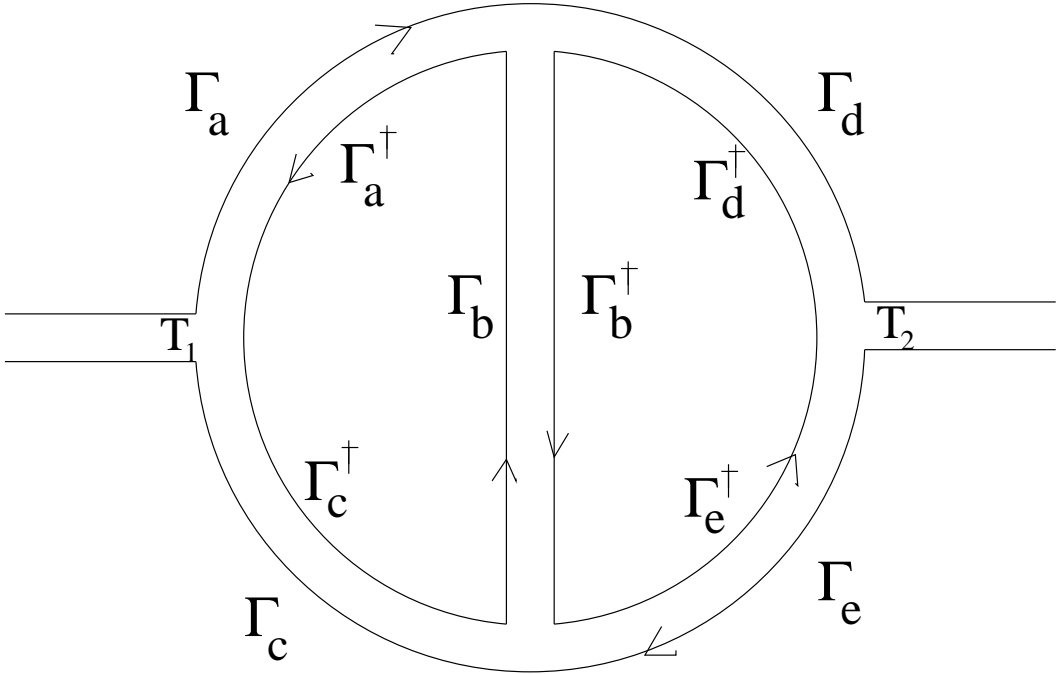}{7in}{3in}\end{center}
\caption{A two loop planar Feynman diagram}
\label{feyn}
\end{figure}  
Consider the planar diagram shown in Fig. (\ref{feyn}). The group theory
factor is
\begin{equation}
g^4\frac{1}{k^5} \sum_{a,b,c,d,e=1}^{k} \Tr[T_1 \Gamma_a  \Gamma_d  T_2 
\Gamma_e \Gamma_c] \Tr[ \Gamma_{c}^{\dagger} \Gamma_b \Gamma_a^{\dagger}]
\Tr[\Gamma_b^{\dagger} \Gamma_e^{\dagger} \Gamma_d^{\dagger}]
\label{diag}
\end{equation}
For simplicity, we are assuming that the orbifold group is embedded trivially
in the global symmetry group. Since the matrices, $\Gamma$ are in the regular
representation, the diagram is zero unless
\[
\Gamma_{c}^{\dagger} \Gamma_b^{\dagger} \Gamma_a^{\dagger}=1
\]
and 
\[
\Gamma_b \Gamma_e^{\dagger} \Gamma_d^{\dagger}=1.
\]
It is then easy to see that (\ref{diag}) becomes
\[
g^4 N^2 \Tr(T_1 T_2).
\]
The same diagram in the parent theory will be proportional to
\[
g^4 N^2 k^2 \Tr(T_1 T_2).
\]
The factor of $(Nk)^2$ comes from summing over $Nk$ particles running
in the two loops. 
The momentum and spin dependence of the same diagram in both theories
are identical so we see that the two diagrams are the same upto rescaling
of couplings. The general proof is given in \cite{bj} and \cite{martin}.

\section{Orbifolds of Seiberg-Witten Theories}
As discussed in the last section, it has been demonstrated that, at least
perturbatively at large $N$, all correlators of the orbifolded
theory are
equivalent to the corresponding correlators 
in the parent theory up to a rescaling of the gauge coupling at some
fixed large scale, e.g. the Planck scale.
In this section we demonstrate that at least one 
aspect of the non-perturbative
behavior of orbifolded and parent theories are related at large $N$, 
namely the gauge 
coupling functions of Seiberg-Witten theories.  In the 't Hooft limit 
$g^2N$ is held fixed with $g\rightarrow 0$.  Then instanton corrections, 
which are
proportional to powers of $e^{-\frac{8\pi^2}{g^2}}$, are generally negligible.
Exceptions which involve $N$th roots of instanton corrections, for example
the gaugino condensate \cite{martin}, are non-vanishing in the 't Hooft limit 
and vary as $e^{-\frac{8\pi^2}{g^2N}}$.  Furthermore, as discussed in 
\cite{Douglas-Shenker}, monopoles which become massless at large $N$ lead
to important nonperturbative effects.  We find a simple relation
between the (inverse) gauge coupling functions of the parent and orbifolded 
theories nonperturbatively for all $N$ and $g$.

For simplicity we study the
case where the curves for the orbifolded and parent theories are
hyperelliptic.  Then the gauge coupling function at generic points in
moduli space, where the gauge group is broken to a product of U(1) factors, is
given by the period matrix of the corresponding curve and is easily expressed
in terms of integrals over cycles of the curve.  The hyperelliptic curve
\cite{ueno,fulton} can be written in the form
\begin{equation}
y^2=f_{2r+2}(x,s_i) , \label{eq:curve} \end{equation}
where the subscript $2r+2$ is the order of the polynomial $f$ in $x$
and $r$ is the genus
of the curve, which for pure ${\cal N}=2$ supersymmetric Yang-Mills theory
is equal to the rank of the corresponding gauge group.  
The moduli $s_i \,(i=1,\dots,r)$ which parameterize the Coulomb branch are the 
vacuum expectation values of the symmetric gauge invariant operators of the 
theory.

In \cite{erich} a prescription was given for generating the curves for 
${\cal N}=1$
supersymmetric SU$(N)^k$ gauge theory with bifundamental chiral multiplets
from those of the ${\cal N}=2$ SU$(kN)$ pure gauge theory by orbifolding 
the ${\cal N}=2$ theory by the Abelian discrete group $Z_k$.  The curve of the 
orbifolded theory, which was obtained by other means in \cite{csaba}, was 
obtained by keeping only
those terms of the parent theory with moduli invariant under the $Z_k$, and 
rescaling $x\rightarrow
x^k$ in the resulting curve.

\begin{figure}\begin{center}
\PSbox{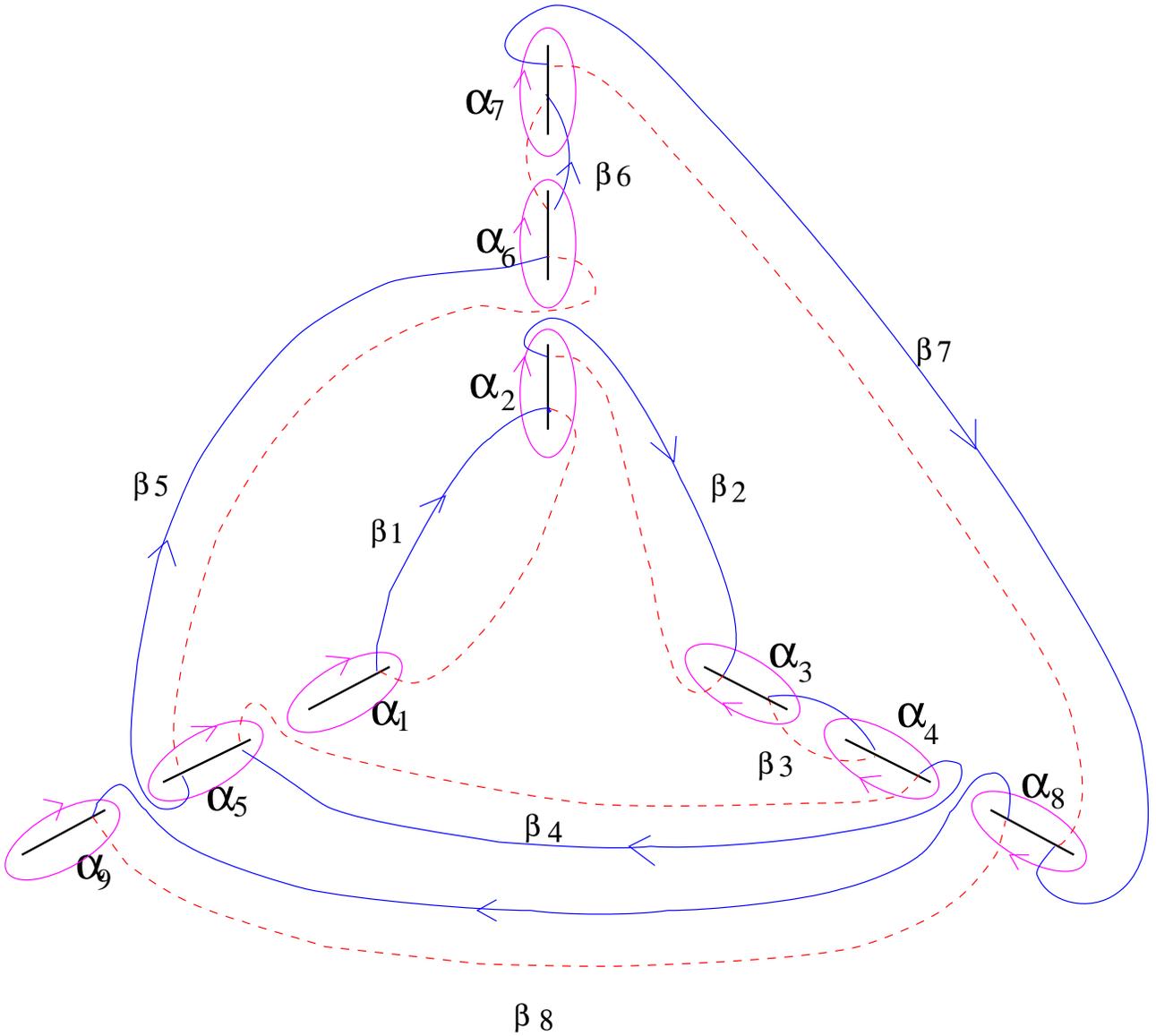}{7in}{7in}\end{center}
\caption{A non-symplectic basis of cycles.  On the orbifold sector of
moduli space the cut $x$ plane is symmetric under the orbifold group, in
this case $Z_3$.}
\label{fig:nonsymplectic}
\end{figure}

\begin{figure}\begin{center}
\PSbox{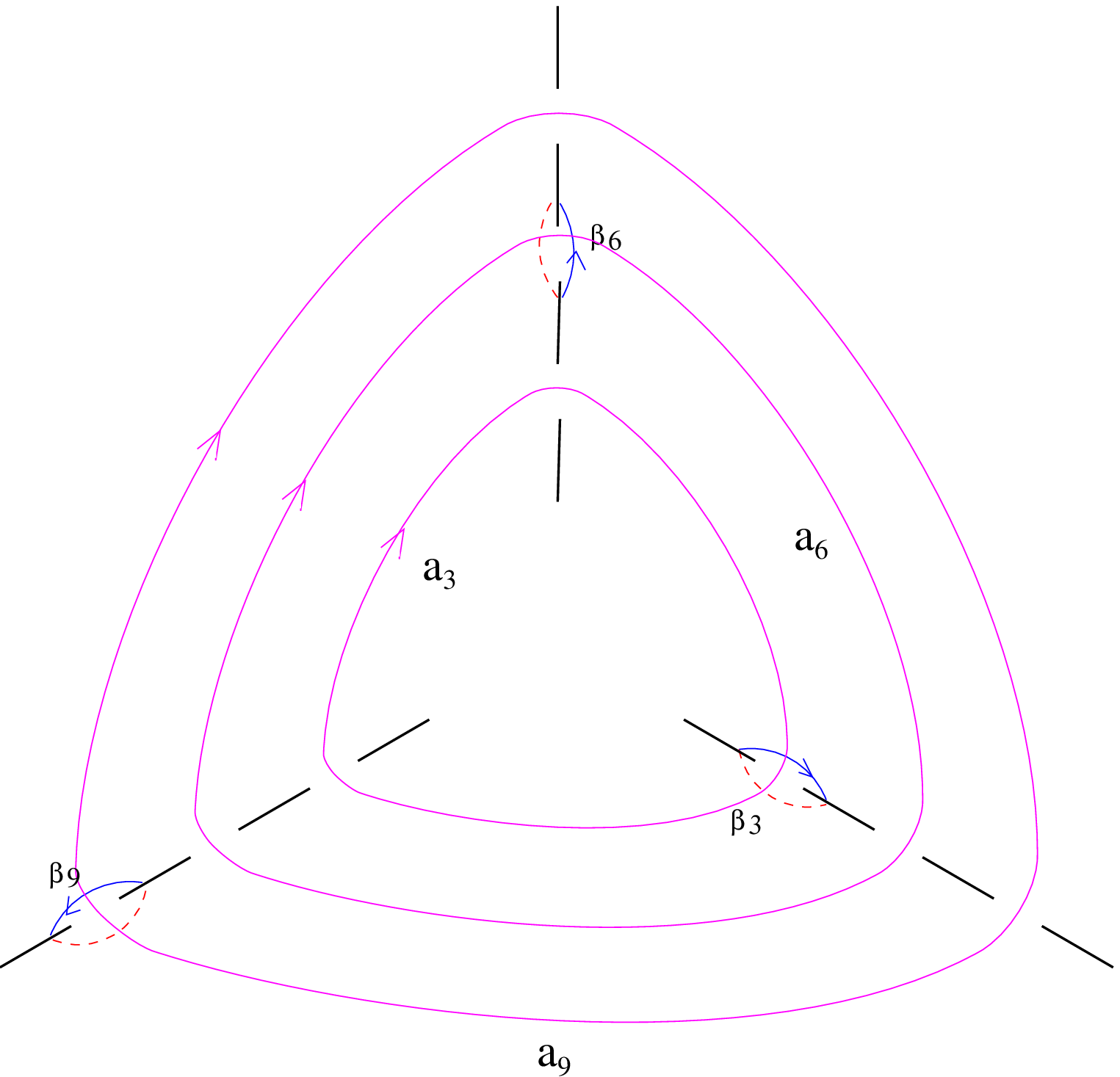}{7in}{7in}\end{center}
\caption{The orbifold sector of a symplectic basis of cycles.}
\label{fig:orbcycles}
\end{figure}

We demonstrate here how the period matrix of the curve corresponding to the
orbifolded theory is related to that of the parent theory.  To be specific
we will consider orbifolding of an SU$(kN)$ gauge theory by $Z_k$.  The genus 
$r$ curve
(\ref{eq:curve}) ($r=kN-1$ here) has $2r$ cycles which are divided into 
$a$ cycles and $b$ cycles with symplectic intersection 
$a_i\cdot b_j=\delta_{ij}$ and $a_i\cdot a_j=b_i\cdot b_j=0$.  
In our case the cuts on the $x$-plane will have a $Z_k$
symmetry, and we label cycles schematically as follows.  We choose a
non-symplectic basis of cycles $(\alpha_i,\beta_i)$ as in 
Fig.~\ref{fig:nonsymplectic}.  Then the appropriate symplectic basis is 
$a_i=\sum_{k=1}^i\alpha_k$, $b_i=\beta_i$.  The subset of
cycles relevant for comparison with the orbifolded theory are shown in
Fig.~\ref{fig:orbcycles}.  They correspond to the $Z_k$ invariant 
$a$ cycles, as discussed below.

The genus $r$ hyperelliptic curve also has a basis of
$r$ holomorphic differentials which can be written \begin{equation}
\omega_j=\frac{x^{j-1}\,dx}{y(x)} \hspace{0.2in} j=1,\dots,r.
\label{eq:omega}
\end{equation}
The matrices of $a$ periods and $b$ periods of the curve are given by 
integrals of the 
differentials (\ref{eq:omega}) over the $a$ and $b$ cycles, respectively,
\begin{equation}
A_{ij}=\int_{a_i}\,\omega_j, \hspace{0.5in} B_{ij}=\int_{b_i}\,\omega_j.
\end{equation}
The period matrix of the curve (\ref{eq:curve}) is then given by
\begin{equation}
\tau_{jl}=B_{jk}A^{-1}_{kl}. \label{eq:tau}
\end{equation}
The identification of the period matrix $\tau_{jl}$ of the hyperelliptic
curve with the gauge coupling function in the ${\cal N}=2$ SU($kN$) theory is 
made via \cite{seiberg-witten,APS,HO} \begin{equation}
\tau_{jl}=\frac{\partial {\bf a}^D_j}{\partial {\bf a}_l},
\label{eq:adual}
\end{equation}
where ${\bf a}_l$ is the vacuum expectation value of the diagonalized adjoint
scalars and ${\bf a}^D_l$ are their duals as described in 
\cite{seiberg-witten}.   Then if for some curve the period matrix 
(\ref{eq:tau}) satisfies
appropriate monodromies around the singularities of
the curve,  then comparing (\ref{eq:tau}) and (\ref{eq:adual}) it
is natural to set \begin{equation}
\frac{\partial{\bf a}^D_j}{\partial s_k}=\int_{b_j}\,\omega_k, \hspace{0.5in}
\frac{\partial{\bf a}_j}{\partial s_k}=\int_{a_j}\,\omega_k.
\end{equation}
Then the orbifold invariant sector of the adjoint VEV's ${\bf a}_j$ 
corresponds to
the invariant sector of $a$ cycles and differentials.  
Alternatively,
the VEV's are given directly by integrals of the Seiberg-Witten differential
\cite{Douglas-Shenker} $\lambda=(1/2\pi i)(x/y)dP(x)$ where $P(x)$ is the
polynomial that appears in the curve $y^2=P(x)^2-\Lambda^{2N}$ of the ${\cal
N}=2$ pure gauge theory.  Since the Seiberg-Witten differential is not
invariant under the $Z_k$ symmetry, only the integrals over the $Z_k$
invariant cycles will be invariant.  Hence, the invariant $a_j$ correspond
to the invariant $a$ cycles.
As matrices, we 
reorganize the periods in a convenient way:
The first $(N-1)\times (N-1)$ block of $A_{jl}$ corresponds to the orbifold 
invariant sector, which we will continue to call $A_{km,kn}$.  

In order to compare corresponding points in the moduli space of the orbifolded
and parent theories, we set all non-invariant moduli in the parent theory
to zero.  We
then study that sector of the gauge coupling functions of the parent theory
that correspond to the U(1) gauge group factors that survive on the moduli
space of the orbifolded
theory.  In the basis of cycles described above, these factors correspond to 
$\tau_{jl}$ for $j$ and $l$ equal to multiples of $k$,
where $k$ is the order of the orbifold group $Z_k$.  It is claimed that this
sector of the inverse period matrix of the curve corresponding to the parent 
theory is
related to the period matrix of the orbifolded theory.  This is verified
as follows.

We can think of the curve $y(x)$ as being defined on a double sheeted cover
of the cut $x$-plane with branch points at the roots of $y(x)$ connected
pairwise to form branch cuts.  On the orbifold sector of the moduli space of 
the parent theory
the curve is a function only of $x^k$, and the roots are of the form 
$x^k=p_i$.  The roots can then be labeled by $p_i$ and a $k$-th root of
unity.  The branch cuts and cycles on the $x$-plane can be distributed
as in Figs.~\ref{fig:nonsymplectic}~and~\ref{fig:orbcycles}.  
The roots of the orbifolded curve are then given by $p_i$.
The holomorphic differentials (\ref{eq:omega}) corresponding to the orbifolded
sector are of the form \begin{equation}
\omega_{kj}=\frac{x^{kj-1}\,dx}{y(x^k)},\hspace{0.2in} j=1,\dots,r.
\label{eq:omega2} \end{equation}
These differentials are invariant under multiplication of $x$ by 
$e^{2m\pi i/k}$, so
the integrals over the $\alpha$ and $\beta$ cycles in 
Fig.~\ref{fig:nonsymplectic} are invariant under similar rotations.
The noninvariant differentials are all multiplied by $e^{2m\pi i/k}$ for some
integer $m$ under multiplication of $x$ by $e^{2\pi i/k}$.  Then the
integrals of the noninvariant differentials over the invariant $a$ cycles
vanish.  For example, suppose $\omega\rightarrow \omega e^{2m\pi i/k}$ when
$x\rightarrow x\,e^{2\pi i/k}$.  Then
\begin{equation}
\int_{a_k}\omega_{{\rm noninv}}=(\sum_{n=0}^{k-1}e^\frac{2mn\pi i}{k})\,
\int_{\alpha_k}\omega_{{\rm noninv}} =0.
\end{equation}

\begin{figure}\begin{center}
\PSbox{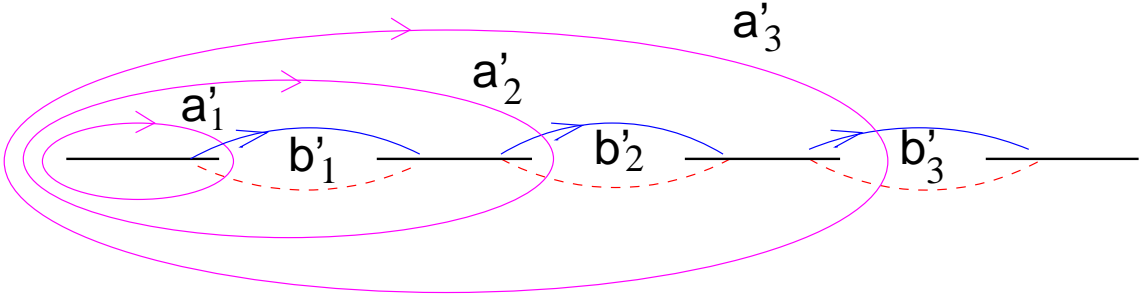}{6.5in}{2in}\end{center}
\caption{Cycles of the orbifolded theory.}
\label{fig:orb}
\end{figure}

Now, consider the invariant sector of the matrix $A_{jl}$ of $a$ periods.  It
is given by \begin{eqnarray}
A_{kj,kl}&=&\int_{a_{kj}} \frac{x^{kl-1}dx}{y(x^k)}
 =\int_{a_{kj}}\frac{\tilde{x}^{l-1}d\tilde{x}}{k\,y(\tilde{x})} \nonumber \\
 &=&\int_{a^\prime_j}\frac{\tilde{x}^{l-1}d\tilde{x}}{y(\tilde{x})} 
\label{eq:Aorb} \\ 
 &=&A^{{\rm orb}}_{jl}, \nonumber
\end{eqnarray}
where in the first line of (\ref{eq:Aorb}) we changed variables $x
\rightarrow \tilde{x}=x^k$, and the cycles $a^\prime_j$ are the $a$ cycles of
the orbifolded theory, drawn schematically in Fig.~\ref{fig:orb}.
An additional factor of $k$ arises in the second line because the invariant
$a$ cycles become a sum over the $a$ cycles of the orbifolded theory $k$ 
times. 
Hence, the orbifold invariant sector of the $a$ periods of the parent theory 
is equivalent to the matrix of $a$ periods of the orbifolded theory.
Hence, the matrix $A_{jl}$ takes the block lower triangular form

\begin{equation}
A=\left(\begin{array}{ccc}
   \left(\begin{array}{ccccccc}
 & & & & &  & \\
\,\, & & & A^{{\rm orb}} & & &\,\, \\
 & & & &   \end{array}\right) & &\left(\begin{array}{ccccccccccc} 
				 0 && 0 && 0 && \cdots && 0 && 0 \\
				 \vdots && \vdots && \vdots && \vdots && 
				   \vdots && \vdots\\
				 0 && 0 && 0 && \cdots && 0 && 0
\end{array}\right)\\
\left(\begin{array}{c}
  \\ \\
\int_{a_{\rm non}}\omega_{{\rm inv}} \\ \\ \\
\end{array}\right) & & \left(\begin{array}{c} \\ \\
 		\int_{a_{\rm non}}\omega_{{\rm noninv}} \\ \\ \\
		\end{array}\right) \end{array}\right).
\end{equation}

A similar analysis applies to the $b$ periods.  By invariant $b$ cycles
we will mean those $b$ cycles that have intersection with the invariant $a$
cycles, and similarly for noninvariant $b$ cycles.  In this case, the integrals
of the invariant differentials over the noninvariant cycles vanish.  The
reason is that the integrals are all of the form \begin{equation}
\int_{p_i^{1/k}}^{\omega p_i^{1/k}}\frac{x^{kj-1}\,dx}{y(x^k)},
\label{eq:bnon} \end{equation}
for some root $p_i$ of $y(x)$.  Then letting $x\rightarrow x^k$ the path of
integration contracts to a point, and the integral (\ref{eq:bnon}) vanishes.

In the basis of cycles and
one forms given above, the invariant sector of $b$ periods of the parent
theory is simply rescaled compared to the $b$ periods of the 
orbifolded theory, \begin{eqnarray}
B_{kj,kl}&=&\int_{b_{kj}} \frac{x^{kl-1}dx}{y(x^k)}
 =\int_{b_{kj}}\frac{\tilde{x}^{l-1}d\tilde{x}}{k\,y(\tilde{x})} \nonumber \\ 
 &=&\int_{b^\prime_j}\frac{\tilde{x}^{l-1}d\tilde{x}}{k\,y(\tilde{x})} 
\label{eq:Borb} \\
 &=&\frac{1}{k}B^{{\rm orb}}_{jl}. \nonumber
\end{eqnarray}
There is no additional factor of $k$ in the second line of (\ref{eq:Borb}) as
in (\ref{eq:Aorb}) since the relevant $b$ cycles of the parent theory 
corresponds to single $b$ cycles of the orbifolded theory and not a multiple
cover as for the $a$ cycles.
Hence, the matrix of $b$ periods takes the block upper triangular form
\begin{equation}
B=\left(\begin{array}{ccc}
   \left(\begin{array}{c}
 \\
 \frac{1}{k}B^{{\rm orb}}  \\ \\
  \end{array}\right) & &\left(\begin{array}{ccc}
 & &  \\
 &\int_{b_{\rm inv}}\omega_{{\rm noninv}}  &\\
 & & \end{array}\right) \\
\left(\begin{array}{ccccc} 
				 0 && \cdots && 0  \\
				 0 && \cdots && 0  \\
				 \vdots &&\vdots  && \vdots  \\
				 0 && \cdots && 0  \end{array}\right) & &
\left(\begin{array}{ccc}
 & &  \\
 &\int_{b_{\rm non}}\omega_{{\rm noninv}}  & \\
 & & \\
 & & \end{array}\right) \end{array}\right).
\end{equation}
The orbifold sector of the inverse (or dual) period matrix, $\tau^{-1}=
AB^{-1}$, of the parent theory is then simply related to that of the orbifolded
theory.  Namely, 
\begin{equation}
\tau^{-1\,{\rm parent}}_{kj,kl}=k\tau^{-1\,{\rm orb}}_{jl}.
\label{eq:finally} \end{equation}
Therefore, the period matrix of the orbifolded theory is determined by that
of the parent theory.

This is the main result of this section.  Equation (\ref{eq:finally}) relates 
the gauge coupling 
functions of the orbifold theory and the orbifold sector of the parent theory. 
It is valid for all $N$ and $g$,
and can be extended to more complicated orbifolds.

\section{Anomaly Positivity Tests}
In a series of papers \cite{josh,anomaly papers} it was demonstrated that in
supersymmetric theories at conformal fixed points there are constraints
on global anomalies which follow from unitarity.  The argument
relies on the supersymmetry multiplet structure which mixes the $R$-charge 
anomaly and the stress tensor trace anomaly, and positivity of central 
functions which appear in the operator product expansion of a product of two
currents.  At conformal fixed points these central functions are central
charges which can be calculated in terms of global anomalies by 't Hooft
anomaly matching.  This procedure is not valid away from a 
fixed point because corrections to the relation between the central functions
and anomalies are proportional to the $\beta$ function 
\cite{josh}.  The argument
can be reversed to give evidence for or against the existence of a conformal
fixed point in a theory.  One assumes that a theory has a fixed point
and calculates the various central charges in terms of 't Hooft matched global 
anomalies.  If
a central charge calculated this way is negative, then the theory could not
have a conformal fixed point.  If the central charge is positive when 
calculated this way, then a strong statement cannot be made, but positivity
provides weak evidence for the existence of a conformal fixed point.

The Weyl anomaly coefficient, which must be positive at an infrared fixed 
point, can be written in terms of the U(1)$_R^3$ and U(1)$_R$ anomalies as 
\begin{equation}
c_{IR}=\frac{1}{32}(9\,U(1)_R^3-5U(1)_R)=\frac{1}{32}\left(4\,{\rm dim}\,G+
\sum_i({\rm dim}\,R_i)(1-r_i)(5-9(1-r_i)^2)\right),
\end{equation}
where dim$\,G$ is the dimension of the gauge group and the sum is over all
representations $R_i$ of matter chiral multiplets in ${\cal N}=1$ language with
$R$-charges $r_i$.
For example, for ${\cal N}=1$ supersymmetric SU($N$) QCD with $N_f$ flavors
of chiral multiplets, the 't Hooft matched $R$ charge of each flavor is 
$\frac{N_f-N}{N_f}$.  The Weyl anomaly coefficient is \cite{josh} 
\begin{equation}
c_{IR}^{SQCD}=\frac{1}{32}\left(4\,(N^2-1)+
2NN_f(\frac{N}{N_f})(5-9(\frac{N}{N_f})^2)\right), \end{equation}
which is easily checked to be positive in the conformal region 
$\frac{3N}{2}<N_f<3N$.  

If the theory has a global flavor symmetry, then the flavor central charge
is required to be positive.  At a conformal fixed point it is given by
\begin{equation}
b_{IR}=-3U(1)_RF^2=3\sum_{ij}({\rm dim}\,R_i)(1-r_i)\mu_i,
\end{equation}
where $\mu_i$ is the Dynkin index of the representation $R_i$.

The Euler anomaly coefficient $a$, which is believed to satisfy
the Zamolodchikov $C$ theorem in four dimensions \cite{josh,cardy,latorre}, 
is also expected to be positive at fixed points.  In addition, the $C$ theorem
requires that the flow of the Euler anomaly be positive:  $a_{UV}-a_{IR}>0$.
The relations between the Euler anomaly and the $R$ current anomalies are
\cite{josh},
\begin{eqnarray}
a_{IR}=\frac{3}{32}&(3U(1)_R^3-U(1)_R)&=\frac{3}{32}\left(2\,{\rm dim}\,G+
\sum_i({\rm dim}\,R_i)(1-r_i)(1-3(1-r_i)^2)\right) \\
&a_{UV}-a_{IR}&=\frac{1}{96}\sum_i({\rm dim}\,R_i)(3r_i-2)^2(5-3r_i).
\end{eqnarray}

Assume a theory that flows to a conformal fixed point in the infrared has a 
unique anomaly free $R$ symmetry.  Then the positivity conditions are 
satisfied in that theory.  If we consider orbifolding that theory, then 
the positivity conditions will remain to be true at large $N$.  This is
the case because the dimension of the orbifolded group at large $N$ is rescaled
by $1/k$, as are the dimensions of each matter representation.  The anomaly
free $R$-charges remain the same in this procedure.  Hence, $c_{IR}$ is 
rescaled by $1/k$ in the orbifolded theory, but otherwise is the same as 
$c_{IR}$ in the parent theory.  Hence, positivity is preserved at large $N$.
That this is true for all theories which are obtained as orbifolds of
theories with conformal fixed points provides evidence that the orbifolded
theories also have conformal fixed points, as expected by the large $N$
orbifold correspondence.  One should note that the anomaly positivity 
conditions of \cite{josh} rely on supersymmetry, so we only consider orbifolds
to supersymmetric theories.  
The anomaly calculations do not rely on a planar diagram expansion,
so this result is valid also away from the 't Hooft limit at large $N$.

In our canonical example, large $N$, ${\cal N}=2$ SU($kN$) pure gauge theory 
orbifolded by $Z_k$ to ${\cal N}=1$ SU$(N)^k$ with bifundamentals, the 
dimension of the parent gauge group is
$(kN)^2$, whereas in the orbifolded theory it is $N^2$ for each SU$(N)$ 
factor, or $kN^2$ total.  Similarly, there are $k$ bifundamental chiral
multiplets in the orbifolded theory, as opposed to one adjoint chiral multiplet
in the parent theory.  The dimension of each of the $k$ bifundamentals is 
$N^2$, so
again the dimension of the representation is rescaled by $1/k$ in the
orbifolded theory compared with the parent theory.  The anomaly free $R$
charges of the adjoint chiral multiplet in the parent theory and the
bifundamental chiral multiplets in the orbifolded theory are 0.

At finite $N$ the anomalies are not simply rescaled by $1/k$, so the above
discussion of preservation of the positivity conditions does not carry through.
This might be related to the problem of additional
U(1)'s that appear in the orbifolded theory which decouple at large $N$, but
not otherwise.  If we naively add one U(1) gaugino for each SU($N$) or SU($kN$)
factor, then the anomalies would be simply rescaled by $1/k$ as for large $N$.
Although positivity in orbifolded theories is not guaranteed as it is for
large $N$, we have surveyed orbifolds of a few theories with duals in the 
conformal regime and have not found violation of any of the positivity 
conditions in any of those orbifolds.  A theory which were to violate 
positivity would
imply a violation of the orbifold correspondence at finite $N$, since then
orbifolds of certain theories with infrared fixed points would not have 
infrared fixed points and correlators of these theories in the infrared
would not match.

In \cite{josh} sufficient conditions on the $R$ charges $r_i$
of the chiral multiplets in an ${\cal N}=1$ theory were given for the various
positivity conditions to be met.  Since orbifolding does not change the $R$
charges, if a parent theory satisfies these sufficient conditions then so will
the orbifolded theory.  The result of \cite{josh} was that in all
renormalizable models studied there the flow $a_{UV}-a_{IR}$ satisfied the 
sufficient
condition for positivity $r_i\leq \sqrt{5}/3$ for all chiral superfields 
$\phi_i$; in all models not requiring an accidental U(1) symmetry for 
unitarity,
$b_{IR}$ and $c_{IR}$ were positive by virtue of $1-\sqrt{5}/3<r_i<1$ for all
$\phi_i$.  Hence
we only need to check positivity for $a_{IR}$, and for $b_{IR}$ and $c_{IR}$
in theories with accidental U(1) symmetry.  What follows is two simple tests 
that we have done.  It would be useful to continue this program by
testing the positivity constraints for orbifolds of other theories.

In the conformal region of ${\cal N}=1$ supersymmetric SU($kN$) gauge theory
with $kN_f$ flavors, $3N_c/2<N_f<3N_c$, the anomaly positivity conditions
were shown to be satisfied in \cite{josh}.  If the theory is orbifolded by
embedding the orbifold group trivially in the U(1)$_R$, then the orbifolded
theory is described by $k$ copies of ${\cal N}=1$ SU($N$) gauge theory with
$N_f$ flavors.  This theory is also in the conformal regime and satisfies the 
anomaly positivity conditions.

\vskip .1in

\centerline{Matter content of Kutasov-Schwimmer models.}

\vskip .1in

\begin{center}
\begin{tabular}{|c||c||c|c|c|}  \hline
~& SU$(N_c)$ & SU$(N_f)_Q$ & SU$(N_f)_{\widetilde{Q}}$ & U$(1)_R$  \\ \hline 
$Q$ & $\Yfund$ & $\Yfund$ &~ & $1-\frac{2N_c}{(k+1)N_f}$ \\
\hline
$\widetilde{Q}$ &$ \overline{\Yfund}$ & ~& 
$\Yfund $& $1-\frac{2N_c}{(k+1)N_f}$ \\
\hline
$X $& adj &~ & ~& $\frac{2}{k+1}$\\
\hline 
\end{tabular}\end{center}

\vskip .1in

The Kutasov-Schwimmer models are given by the matter content and charges
in the table above.  We have taken the 
superpotential to be $W={\rm Tr}\,X^3$, where $X$ is the adjoint chiral
superfield.  An orbifold preserving the ${\cal N}=1$
supersymmetry of those models is obtained by embedding the orbifold group in 
the gauge and global symmetries
as for the SQCD case described above, and then the adjoint chiral multiplet
decomposes into an adjoint under each of the SU($N$) factors of the orbifold
theory.  By explicit calculation we find that for all $N_f$ and $N_c$ in the
conformal region without accidental symmetry, $N_c<N_f<2N_c$, 
the anomaly positivity conditions are satisfied.

These results provide evidence that orbifolds of theories with fixed points 
have
fixed points themselves, also hinting at a correspondence between certain 
theories and their orbifolds, even at finite $N$.  A more complete study 
would be useful.  Comments on the ADS/CFT correspondence at finite $N$ were
made recently in \cite{witten}.

\section{Conclusions}
We have demonstrated a simple relation between the gauge coupling functions
of ${\cal N}=2$ SU($kN$) pure gauge theory and ${\cal N}=1$ SU$(N)^k$ gauge
theory with bifundamental chiral multiplets.  If the prescription given in
\cite{erich} is generic for producing Seiberg-Witten curves of orbifolded 
field theories
with a Coulomb branch, then this result is valid for all such theories.  The
problem of anomalous U(1)'s has not been satisfactorily understood in generic
orbifolds of field theories.  In the case studied above, the problem of
anomalous U(1)'s is compensated for by axions in the twisted sectors of the
orbifold theory \cite{erich}.  In that case, the curves obtained are those of 
the orbifolded theory
without the additional U(1)'s or twisted sector fields, as derived in 
\cite{csaba}.  It is not known to us whether this behavior is generic.
This demonstrates a correspondence between one aspect of the 
orbifolded and parent theories at finite $N$.  It is not know to us whether
the Kahler potentials for the two theories behaves similarly.

We studied the anomaly positivity constraints \cite{josh} on theories with 
infrared fixed
points and found that at large $N$ these constraints are satisfied also in
field theory orbifolds of such theories.  At finite $N$ we studied two
classes of theories and their orbifolds and found no violation of the
positivity constraints in the orbifolded theories.  This provides some evidence
that orbifolds of theories with conformal fixed points have conformal fixed
points themselves, in keeping with the parent/orbifold correspondence.

\section*{Acknowledgments}

We are grateful to Dan Freedman, Martin Gremm, Ami Hanany, Lisa Randall and 
Martin Schmaltz for useful conservations. This research was supported in part by the
U.S. Department of Energy under cooperative agreement \# DE-FC02-94ER40818.

\end{document}